\documentclass[prl,aps,reprint,superscriptaddress,amsmath]{revtex4-1}

\usepackage{braket}
\usepackage{hyperref}
\usepackage{graphicx}

\begin{document}

\title{Controlling the dipole blockade and ionization rate of Rydberg atoms \\
in strong electric fields}

\author{Markus Stecker}
\affiliation{Center for Quantum Science, Physikalisches Institut, Eberhard Karls Universit\"at T\"ubingen, Auf der Morgenstelle 14, D-72076 T\"ubingen, Germany} 

\author{Raphael Nold}
\affiliation{Center for Quantum Science, Physikalisches Institut, Eberhard Karls Universit\"at T\"ubingen, Auf der Morgenstelle 14, D-72076 T\"ubingen, Germany} 

\author{Lea-Marina Steinert}
\affiliation{Center for Quantum Science, Physikalisches Institut, Eberhard Karls Universit\"at T\"ubingen, Auf der Morgenstelle 14, D-72076 T\"ubingen, Germany} 

\author{Jens Grimmel}
\affiliation{Center for Quantum Science, Physikalisches Institut, Eberhard Karls Universit\"at T\"ubingen, Auf der Morgenstelle 14, D-72076 T\"ubingen, Germany} 

\author{David Petrosyan}
\affiliation{Center for Quantum Science, Physikalisches Institut, Eberhard Karls Universit\"at T\"ubingen, Auf der Morgenstelle 14, D-72076 T\"ubingen, Germany}\affiliation{Institute of Electronic Structure and Laser, FORTH, GR-71110 Heraklion, Crete, Greece} 

\author{J\'{o}zsef Fort\'{a}gh}
\affiliation{Center for Quantum Science, Physikalisches Institut, Eberhard Karls Universit\"at T\"ubingen, Auf der Morgenstelle 14, D-72076 T\"ubingen, Germany} 

\author{Andreas G\"unther}
\email[]{a.guenther@uni-tuebingen.de}
\affiliation{Center for Quantum Science, Physikalisches Institut, Eberhard Karls Universit\"at T\"ubingen, Auf der Morgenstelle 14, D-72076 T\"ubingen, Germany}

\date{\today}

\begin{abstract}
We study a novel regime of the Rydberg excitation blockade using highly 
Stark-shifted, yet long-living, states of Rb atoms subject to electric 
fields above the classical ionization limit. Such states allow tuning the 
dipole-dipole interaction strength while their ionization rate can be changed
over two orders of magnitude by small variations of the electric field. 
We demonstrate laser excitation of the interacting Rydberg states followed
by their detection using controlled ionization and magnified imaging 
with high spatial and temporal resolution.
Our work reveals the hitherto unexplored possibilities to control 
the interaction strength and dynamically tune the ionization and detection 
of Rydberg atoms, which can be useful for realizing and assessing 
quantum simulators that vary in space and time.
\end{abstract}

\maketitle

Atoms excited to highly polarizable Rydberg states can exhibit strong
interactions and long-range correlations, which makes them promising systems 
for quantum information processing \cite{Jaksch2000,Lukin2001,Saffman2010} 
and quantum simulations \cite{Weimer2010}.
The interatomic interactions can be of dipole-dipole or van der Waals 
type \cite{Reinhard2007}, often manifested in the experiments through the 
so-called Rydberg blockade, i.e., suppression of laser excitation of more 
than one Rydberg atom within a certain blockade volume. 

The efficiency of the blockade and the fidelity of blockade-based 
quantum gates depend on the strength of the Rydberg-Rydberg interactions 
\cite{Saffman2005,Zhang2012,Petrosyan2017}. A tunable interaction potential 
is also highly desirable for quantum simulators  \cite{Weimer2010}. 
Following the first demonstrations of the Rydberg blockade \cite{Tong2004,Singer2004}, 
much effort has been devoted to enhancement and tuning of the interatomic interactions, 
e.g., by shifting the atomic Rydberg states with a weak electric field to 
a F\"orster resonance \cite{Vogt2006,Reinhard2008a, Ravets2014}, 
using states with near resonant dipole-dipole interactions at zero field \cite{Reinhard2008}, 
employing the AC Stark shifts of the Rydberg states \cite{Bohlouli-Zanjani2007}, 
and using rotary echo \cite{Thaicharoen2017} or weak static electric fields 
\cite{Vogt2007, Thaicharoen2016,Goncalves2016}. 
Rydberg blockade between two individual atoms \cite{Urban2008,Gaetan2009} and 
two-qubit gates and entanglement have been achieved 
\cite{Isenhower2010,Wilk2010, Maller2015,Levine2018}, and single collective Rydberg excitations 
in blockaded atomic ensembles have been observed \cite{Weber2015,Ebert2015,Zeiher2015}. 
Finally, simulations of quantum Ising models with atoms in a lattice or arrays of microtraps 
have been demonstrated \cite{Schauss2012,Schauss2015,Labuhn2016,Bernien2017,Lienhard2018,Guardado-Sanchez2018}.
Spatially resolved measurement of the Rydberg blockade has been studied 
with separate excitation areas \cite{VanDitzhuijzen2008, Carroll2004}, 
quantum gas microscopes \cite{Schauss2012,Schauss2015}, 
interaction enhanced imaging \cite{Gunter2012,Gunter2013} and direct imaging 
by field ionization \cite{Schwarzkopf2011,Schwarzkopf2013,Thaicharoen2015,Fahey2015}. 

In this work, we investigate Rydberg-Rydberg interactions of atoms in a strong static 
electric field close to the classical ionization limit, which is usually deemed 
inapplicable for such experiments. Yet, we identify highly Stark-shifted states with 
tunable dipole moments and low ionization rates suitable for observing correlations 
stemming from the Rydberg blockade. We vary the interatomic interaction strength by 
small changes in the external electric field and observe Rydberg blockade of different 
strength and range. Furthermore, we demonstrate a detection scheme in which the atoms
in a long-living Rydberg state are adiabatically transferred to a state with a strongly
enhanced ionization rate. Spatially resolved detection is then achieved by imaging 
the ions with a high resolution ion microscope \cite{Stecker2017}. 

\begin{figure}[tbp]
\includegraphics[width=8.7cm]{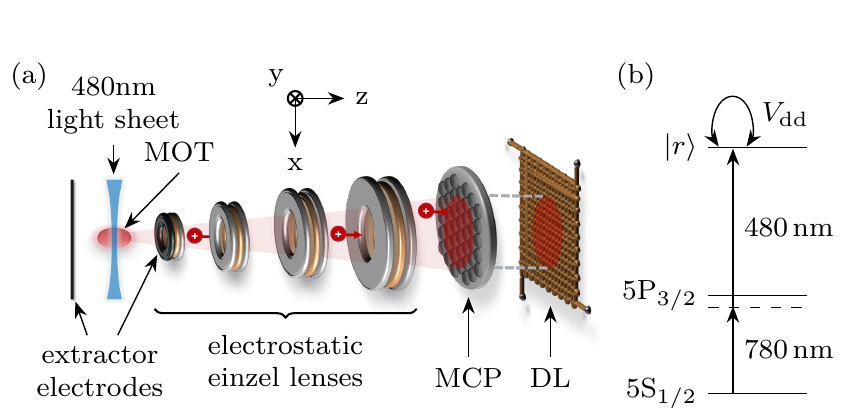}
\caption{ (a) Experimental setup for Rydberg excitation of atoms followed by spatially 
resolved ion detection. Atoms in a magneto-optical trap (MOT) are excited 
to the Rydberg state in the overlap region of the 780\,nm MOT laser and the 480\,nm laser light sheet. 
Field ionized Rydberg atoms are imaged via electrostatic einzel lenses onto a multi-channel plate (MCP) 
and electronically detected via a delay line anode (DL) with single ion sensitivity and high temporal 
and spatial resolution. 
(b) Rydberg excitation scheme for $^{87}$Rb atoms with a two-photon transition from the $5S_{1/2}$ ground state 
to a high lying Rydberg state $\ket{r}$ via the intermediate $5P_{3/2}$ state.
The atoms in state $\ket{r}$ interact with each other via the dipole-dipole potential $V_{\mathrm{dd}}$.}
\label{fig:Setup}
\end{figure}

Our experimental setup is illustrated in Fig. \ref{fig:Setup}(a). 
It consists of a vacuum chamber with a standard magneto-optical trap (MOT) 
for $^{87}\mathrm{Rb}$ atoms and an ion microscope. 
The MOT contains up to $10^7$ atoms at temperature $T \simeq 150\,\mu$K 
and peak density $D \simeq 4 \times 10^{10}\:$cm$^{-3}$. 
Two extractor electrodes can generate the desired electric field at the position of the atoms in the MOT. 
The atoms are excited to the Rydberg state by a two-photon transition from the $5S_{1/2}$ 
ground state via the near-resonant intermediate $5P_{3/2}$ state [see Fig. \ref{fig:Setup}(b)]. 
The lower transition at a wavelength around 780\,nm is provided by the continuous MOT cooling laser, 
while the upper transition at around 480\,nm is driven by a pulsed laser stabilized 
with a wavelength meter (HighFinesse WS8-2).
This blue laser beam is directed along $x$, perpendicular to the electric field of the extractor 
electrodes and the optical axis of the ion imaging, and is focused to a light sheet with beam waists 
of $w_z \simeq 4.5\,\mu$m and $w_y \simeq 40\,\mu$m in order to limit the excitation depth 
in the $z$ direction of the ion imaging. 
The spatially resolved detection of ions outside of the MOT is performed by an ion microscope 
composed of four consecutive electrostatic lenses and a multi-channel plate (MCP) 
in conjunction with a delay line anode for electronic readout of the individual ion positions 
and time stamps [see Fig. \ref{fig:Setup}(a)]. 
The magnification of the ion microscope is around 1000, resulting in an imaging area 
of 40\,$\mu$m diameter; further details of the experimental setup can be found in \cite{Stecker2017}. 

\begin{figure}[tbp]
\includegraphics[width=8.7cm]{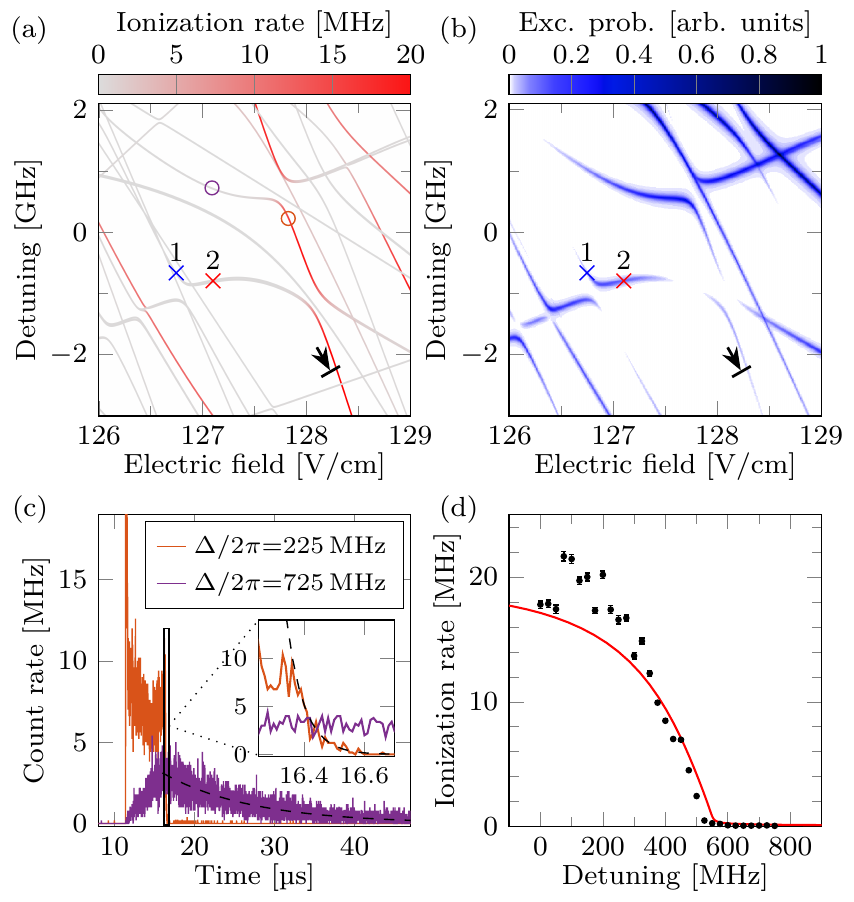}
\caption{(a) Numerically calculated Stark spectrum of $^{87}\mathrm{Rb}$ atoms near the classical ionization threshold at 127.2\,V/cm, with the detuning relative to the unperturbed $43S_{1/2}$ state.
Ionization rates of the individual Stark lines range from non-ionizing (gray) to strongly ionizing (red).
Crosses indicate the strongly (1) and weakly (2) interacting resonances used in the Rydberg blockade 
experiment of Fig. \ref{fig:Blockade}. The arrowhead with a delimiter mark indicates 
the ionizing state that the Rydberg state is transferred to for the blockade measurements.
(b) Excitation strength of Rydberg resonances calculated for the laser polarizations 
and the dipole matrix elements of the corresponding optical transitions,
starting from the $5S_{1/2}$ ground state [see Fig. \ref{fig:Setup}(b)].
(c)~Examples of the ion signal decay, for the two resonances indicated by open circles in panel (a),
for the extraction of Rydberg state ionization rates.
(d) Measured (black dots) and calculated (solid red line) ionization rates along the Stark 
line with open circles in panel (a).}
\label{fig:cap_theory}
\end{figure}

The energy landscape of highly Stark-shifted Rydberg atoms exhibits an intricate form \cite{Grimmel2015} 
and is governed by strongly varying line broadenings due to ionization. 
From our numerical calculations \cite{Grimmel2017} using a complex absorbing potential (CAP) 
\cite{Kosloff1986,Riss1993,Sahoo2000}, we identify several states with small ionization rates 
and different dipole moments. 
In Fig.~\ref{fig:cap_theory}(a) we show the calculated Stark spectrum and ionization rates
around the (unperturbed) $43S_{1/2}$ state of $^{87}\mathrm{Rb}$ with the electric field 
in the interval of 126-129\,V/cm around the classical ionization threshold at 127.2\,V/cm. 
We choose this field region because it features multiple resonance lines that undergo a strong change 
in ionization rate, from long-living to strongly ionizing, within a small interval of the electric field. 
The numerically calculated laser excitation spectrum of the Stark-shifted Rydberg states 
is shown in Fig. \ref{fig:cap_theory}(b).  

To experimentally verify the theoretically predicted ionization rates, we excite the atoms to the Rydberg 
states lying on one of the Stark lines at different electric fields, with the corresponding detuning 
of the blue laser, and simultaneously detect the arrival times of the ions reaching the detector. 
With the MOT lasers continuously on, the blue laser is pulsed for 1-5\,$\mu$s, each pulse typically yielding 
1-30 ions, depending on the transition strength and pulse duration. We repeat the excitation pulse 
several thousand times to derive a histogram of the ion arrival times. We then determine the ionization 
rate of the state from the decay of the ion signal after the end of the excitation pulse, as illustrated
in Fig. \ref{fig:cap_theory}(c). 
We note that radiative decay to lower states \cite{Gallagher1994} and population redistribution 
by black body radiation \cite{Beterov2009} can alter the ionization 
rate of the Rydberg atoms. Yet, we do not expect significant contributions of these processes, having typically 
rates of a few tens of kHz, as compared to our observed ionization rates in the MHz range. 
In the strongly Stark-shifted regime, however, the spontaneous and black body induced transition rates 
may differ from their zero-field values. Nevertheless, the experimental data is in good agreement 
with the results of the numerical calculation, as seen in Fig. \ref{fig:cap_theory}(d). 
Deviations between theory and experiment arise from the fact that the free parameter of the CAP potential 
was optimized for a broad spectral range and not for a single resonance \cite{Grimmel2017}.

Our results thus demonstrate that by a small variation of the electric field the ionization rate 
of the atomic Rydberg states can be precisely tuned in a wide interval, from a few tens of kHz 
up to several MHz, which can facilitate the temporally and spatially resolved detection of Rydberg atoms. 
We can excite the atoms to a long-living Rydberg state and then adiabatically transfer them to a state with well defined ionization rate by a small change of the electric field.  
Typically, we change the electric field by about 1\,V/cm within 1$\mu$s. 
Traditionally, Rydberg atoms are ionized by simply switching on, or ramping up,
the electric field to a much higher value. In contrast, our approach permits a high degree of control 
of ionization. We can dynamically tune the ionization rate to be, on the one hand, strong enough for 
the atoms not to move or decay significantly while they are ionized, and, on the other hand, 
low enough to avoid saturation of the ion detector. Moreover, our method permits the reversal 
of the ionization rate from strong to weak by an adiabatic change of the electric field. 
This is virtually impossible when ionizing by switching on a strong field which results 
in rapid redistribution of the population of the initial Rydberg state over many Stark-shifted 
states \cite{Feynman2015}.

We now turn to the study of Rydberg blockade in the atomic ensemble. 
We choose the electric field values and the corresponding frequency of the blue laser 
to resonantly excite slowly ionizing Rydberg states characterized by different
dipole moments and thereby different strengths of the interatomic interactions. 
Specifically, we selectively address a single Stark line at two positions, 1 and 2,
as indicated in Fig. \ref{fig:cap_theory}(a,b). Resonance 1 corresponds to the
electric field of $F = 126.75\,$V/cm and a laser detuning $\Delta /2\pi = -664\,$MHz
(relative to the unperturbed $43S_{1/2}$ state), while resonance 2 corresponds 
to $F=127.1\,$V/cm and $\Delta /2\pi =-794\,$MHz.
The dipole moment $p_z$ of a Stark-shifted state with energy $E(F)$ in an electric field $F$ 
can be determined from the slope of the resonance line $p_z =-dE(F)/dF$. 
For the first resonance with the large energy slope we obtain the dipole moment 
$p_z \simeq 2000ea_0 =1.69 \times 10^{-26}$\,Cm, which should lead to a strong dipole-dipole 
interaction with the coefficient $C_3/2\pi \simeq 3.9\,\textrm{GHz}\,\mu\textrm{m}^3$ (see below)
and thereby well-pronounced blockade. 
In contrast, the small slope around the second resonance results in a smaller dipole 
$p_z \simeq -210ea_0 = -1.76 \times 10^{-27}$\,Cm and an interaction coefficient 
$C_3/2\pi \simeq 42\,\textrm{MHz}\,\mu\textrm{m}^3$,
leading to much weaker excitation blockade.

For the spatially resolved detection of Rydberg atoms, the blue laser pulse excites the atoms 
to slowly ionizing Rydberg states. We set the duration of the pulse to $\tau = 5\,\mu$s 
for resonance 1 with weaker two-photon excitation strength, and 
to $\tau =1\,\mu$s for resonance 2 with stronger excitation strength [see Fig.~\ref{fig:cap_theory}(b)].  
Following the excitation, the voltage at the extractor electrodes 
is ramped up to about $F = 128.3\,$V/cm within 1$\mu$s using a high voltage switch. 
This transfers the Rydberg atoms to a rapidly ionizing state, as indicated by the arrowhead 
marker in Fig.~\ref{fig:cap_theory}(a). The ions are detected by our ion optics setup which records
the time of flight and positions of the ions hitting the detector. The magnification of the ion-optical 
system is set to 1129. Each excitation-detection cycle typically yields 2-8 ions and is repeated up 
to $3\times 10^5$ times to accumulate good statistics. 

\begin{figure}[tbp]
\includegraphics[width=8.7cm]{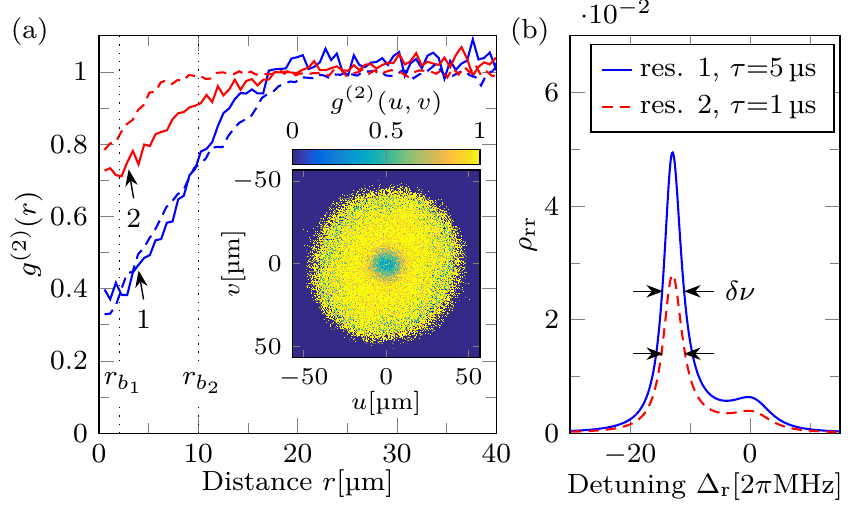}
\caption{(a) Correlation function $g^{(2)}(r)$ versus distance $r$, as obtained from the experimental 
ion measurements for the Rydberg excitation of resonances 1 (thick solid blue line) and 2 (thick solid red line).  The corresponding results from the numerical Monte Carlo simulations are plotted with dashed lines of the same color. The inset shows the experimental two-dimensional correlation 
function $g^{(2)}(u,v)$ for resonance 1. 
(b) Excitation probability $\rho_{rr}$ of Rydberg state $\ket{r}$ for a three-level atom versus 
detuning $\Delta_r$ of the blue laser, as obtained from numerical solution of the density matrix 
equations with the following parameters: 
The cw red laser acts on the lower atomic transition $5S_{1/2} \leftrightarrow 5P_{3/2}$ with 
Rabi frequency $\Omega_{780}/2\pi = 7.4\,$MHz and detuning  $\Delta_{780}/2\pi = 12\,$MHz, while
the blue laser acting on the upper transition  $5\mathrm{P}_{3/2} \leftrightarrow \ket{r}$ with Rabi 
frequency $\Omega_{480}/2\pi = 0.3,0.5\,$MHz is pulsed for time $\tau = 5,1\,\mu$s for resonance 1,2,
respectively. Decay rates of the intermediate and Rydberg states are $\Gamma_{5P}/2\pi = 6\,$MHz
and $\Gamma_{r}/2\pi = 30\,$kHz, and the dephasing rates for the corresponding transitions are
$\gamma_{5P}/2\pi = 0.55\,$MHz and $\gamma_{r}/2\pi = 1\,$MHz. 
The Rydberg excitation linewidth is $\delta \nu \simeq 2\pi \times 3.8\,$MHz for both resonances 1,2.}
\label{fig:Blockade}
\end{figure}

After each excitation and spatially-resolved ion detection, we calculate the two-dimensional 
second order correlation function 
\begin{align}
g^{(2)}(u,v) = \frac{\langle\langle f(x+u,y+v)f(x,y)\rangle\rangle_{x,y}}
{\langle\langle f(x+u,y+v)\rangle\rangle_{x,y}\langle\langle f(x,y)\rangle\rangle_{x,y}},
\end{align}
where $u,v$ are the displacements, 
$f(x,y)=\sum_{i=1}^N \delta(x-x_i)\delta(y-y_i)$ is the detector function with the coordinates 
$x_i$ and $y_i$ of the $N$ incoming ions on the detector, and $\langle\langle \cdot \rangle\rangle_{x,y}$
denotes the average over the $x$ and $y$ positions. The correlation results for the individual cycles
are summed up. In order to compensate for the spatially inhomogeneous excitation probability 
(due to inhomogeneous laser power) and the finite detector size, the result is normalized to 
the correlation function calculated from all events at once. In the inset of Fig.~\ref{fig:Blockade}(a)
we show an example of such a measured correlation function for the excitation of  
the strongly-interacting Rydberg state. The correlation function has a reduced value $g^{(2)}(u,v) < 1$
in a certain blockade region $\sqrt{u^2+v^2} < r_b \sim 10\,\mu$m around the origin ($u,v=0$) 
with no apparent anisotropy. Hence, in the vicinity of a Rydberg excited atom, the probability 
of another excitation is strongly suppressed, which is a clear manifestation of the Rydberg blockade.

In Fig.~\ref{fig:Blockade}(a) we show the correlation function $g^{(2)}(r)$ versus distance $r$ between 
the Rydberg excitations as obtained by radial binning and angle averaging the two-dimensional correlation 
function $g^{(2)}(u,v)$. At large distances, $r > r_b$, there are no correlations between the Rydberg
excitations, corresponding to $g^{(2)}(r) \simeq 1$. At smaller distances, $r \lesssim r_b$, the suppression 
of multiple Rydberg excitations, $g^{(2)}(r) < 1$, signifies the blockade correlations. Clearly the blockade 
strength and distance are much larger for the strongly-interacting Rydberg state (resonance 1) than 
for the weakly interacting state (resonance 2).

The dipole-dipole interaction potential between pairs of atoms in the Rydberg state 
with a permanent dipole moment $p_z$ along the $\hat{z}$ direction is given by
\begin{equation}
V_{\mathrm{dd}} = \hbar \frac{C_3}{r^3} (1 - 3 \cos^2 \theta), \label{eq:Vdd}
\end{equation}
where $C_3 \equiv p_z^2/(4 \pi \epsilon_0 \hbar)$ is the interaction coefficient referred to above, 
$r \equiv |\boldsymbol{r}|$ is the distance between the atoms, and $\theta$ is the angle between 
vectors $\hat{z}$ and $\boldsymbol{r}$. We can estimate the blockade distance $r_b$ as follows.
The excitation volume is determined by the blue laser field which is a thin light sheet, 
$w_z \ll w_y$, oriented perpendicular to the strong electric field $F$ of the ion extractor 
electrodes in the $\hat{z}$ direction. We may therefore assume $\theta = \pi/2$ in Eq.~(\ref{eq:Vdd}). 
The blockade distance is then defined via $V_{\mathrm{dd}}/\hbar \geq \delta \nu$ as the distance 
$r_b = (C_3/\delta \nu)^{1/3}$ at which the interaction-induced level shift of the Rydberg 
state $\ket{r}$ is equal to, or larger than, the excitation linewidth $\delta \nu$ of 
the atoms subject to near-resonant lasers. 
In Fig.~\ref{fig:Blockade}(b) we show the spectra of Rydberg excitation of three-level atoms 
obtained by numerically solving the density matrix equations for the stated parameters. For both resonances 1 and 2, we
obtain the Rydberg excitation linewidth $\delta \nu \simeq 2\pi \times 3.8\,$MHz,
leading to $r_b \simeq 10\,\mu$m for resonance 1, and $r_b \simeq 2.2\,\mu$m 
for resonance 2.    

In Fig.~\ref{fig:Blockade}(a) we compare the experimentally obtained correlations 
between the Rydberg excitations with the results of our numerical simulations. 
In the simulations, we place the atoms with the average density $D \simeq 0.04\:\mu$m$^{-3}$ 
at random positions in a sufficiently large volume. 
The three-level atoms are driven by a spatially and temporally uniform red laser 
acting on the lower transition, and a pulsed blue laser, focused to a light sheet, acting 
on the upper transition, with the atomic parameters as in Fig.~\ref{fig:Blockade}(b). 
The laser fields are in two-photon resonance with an unperturbed Rydberg level of the atoms,
but the Rydberg level of each atom is shifted by the interaction with all the other Rydberg
excited atoms, which translates into the corresponding detuning $\Delta_r$ of the laser. 
The Rydberg excitations at different positions are then obtained by Monte Carlo 
sampling of the excitation probabilities of the atoms \cite{Ates2007,Petrosyan2013a,Petrosyan2013b}.
After the excitation pulse, the Rydberg state atoms are assumed ionized and detected in the $x,y$ 
plane by the ion detector. We take the ion detection efficiency $\eta = 0.7$ and a Gaussian position 
uncertainty with standard deviation $\sigma_{x,y} = 2.3\,\mu$m for resonance 1 and 
$\sigma_{x,y} = 1.2\,\mu$m for resonance 2. The larger position uncertainty for the former
case is intended to model the thermal atomic motion during the longer $\tau = 5\,\mu$s
excitation time. The correlation functions shown in Fig.~\ref{fig:Blockade}(a) are obtained
after averaging over $10^5$ independent realizations of the dynamics of Rydberg excitations 
and detection in random atomic ensembles. The numerically obtained correlation functions
are in reasonably good agreement with the corresponding experimental ones.

Note, that even for the strongly-interacting case of resonance 1, the correlation function $g^{(2)}(r)$ 
does not reach zero for $r \to 0$. This is due to the finite thickness $w_z$ of the excitation volume
where the angular dependence and zeroes of the dipole-dipole potential $V_{\mathrm{dd}}$ 
partially attenuate the blockade. But even in perfectly 1D or 2D geometries, 
the Rydberg blockade due to the dipole-dipole potential $V_{\mathrm{dd}} \propto r^{-3}$ 
is much ``softer'', and the blockade distance is more uncertain, as compared to 
those for the van der Waals potential $V_{\mathrm{vdW}} \propto r^{-6}$ \cite{Petrosyan2013b}. 

To conclude, we have shown that long-living Rydberg excitations and strong and tunable interatomic 
interactions can be observed in highly Stark-shifted states of atoms in strong electric fields. 
In many experiments, the atoms are subject to strong or residual electric fields, especially in chip-based quantum systems, where adsorbates on the surfaces play 
a crucial role \cite{McGuirk2004,Obrecht2007,Tauschinsky2010,Hattermann2012}.
Such experiments often aim at strongly coupling Rydberg atoms to surface-based quantum systems or superconducting coplanar waveguide resonators \cite{Petrosyan2009,Hogan2012,Sarkany2015,Sarkany2018,Petrosyan2019}, requiring the atoms to be brought close to the surface \cite{Bernon2013,Hattermann2017}.
Then, the fields typically have to be compensated by a set of electrodes to encompass all spatial
directions, which can be challenging and cumbersome. Instead of striving to attain a zero field 
condition, we have shown in this work that the complicated ``spaghetti region" of the Stark spectrum with
its diversity of states can be seen as an advantage rather than annoyance, allowing for tailoring 
excitation, interaction and ionization of Rydberg states.

\begin{acknowledgments}
This work was supported by the Deutsche Forschungsgemeinschaft through SPP 1929 (GiRyd). 
M.S. acknowledges financial support from Landesgraduiertenf\"orderung Baden-W\"urttemberg. 
\end{acknowledgments}

\end{document}